\pdfoutput=1
\documentclass[11pt]{article}

\usepackage{ACL2023}

\usepackage{times}
\usepackage{latexsym}

\usepackage[T1]{fontenc}
\usepackage[utf8]{inputenc}

\usepackage{microtype}

\usepackage{inconsolata}

\usepackage{amsmath,amsfonts,bm}

\def\eqref#1{equation~\ref{#1}}
\def\1{\bm{1}}

\def\ve{{\bm{e}}}

\def\vh{{\bm{h}}}

\def\vx{{\bm{x}}}

\def\mE{{\bm{E}}}

\def\mW{{\bm{W}}}
\def\mX{{\bm{X}}}

\DeclareMathAlphabet{\mathsfit}{\encodingdefault}{\sfdefault}{m}{sl}
\SetMathAlphabet{\mathsfit}{bold}{\encodingdefault}{\sfdefault}{bx}{n}

\newcommand{\R}{\mathbb{R}}

\newcommand{\softmax}{\mathrm{softmax}}

\usepackage{hyperref}
\usepackage{url}
\usepackage{graphicx}
\usepackage{booktabs}
\usepackage{bbding}
\usepackage{multirow}
\usepackage{array}
\usepackage{stfloats}

\newcommand{\textblue}[1]{\textcolor{blue}{\textbf{#1}}}

\title{Hybrid and Collaborative Passage Reranking}

\author{Zongmeng Zhang$^{1}$, Wengang Zhou$^{2}$, Jiaxin Shi$^{3}$, Houqiang Li$^{2}$ \\
		$^{1,2}$University of Science and Technology of China \\
		$^{3}$Huawei Cloud Computing Technologies Co., Ltd. \\
		$^{1}$\texttt{zhangzm@mail.ustc.edu.cn, $^{2}$\{zhwg, lihq\}@ustc.edu.cn, $^{3}$shijx12@gmail.com} \\}

\begin{document}
\maketitle
\begin{abstract}
In passage retrieval system, the initial passage retrieval results may be unsatisfactory, which can be refined by a reranking scheme.  
Existing solutions to passage reranking focus on enriching the interaction between query and each passage separately, neglecting the context among the top-ranked passages in the initial retrieval list. 
To tackle this problem, we propose a \textit{\textbf{Hyb}rid and Collaborative Passage Re\textbf{rank}ing} (\textbf{HybRank}) method, which leverages the substantial similarity measurements of upstream retrievers for passage collaboration and incorporates the lexical and semantic properties of sparse and dense retrievers for reranking. 
Besides, built on off-the-shelf retriever features, HybRank is a plug-in reranker capable of enhancing arbitrary passage lists including previously reranked ones.
Extensive experiments demonstrate the stable improvements of performance over prevalent retrieval and reranking methods, and verify the effectiveness of the core components of HybRank.\footnote{Our code is available at \url{https://github.com/zmzhang2000/HybRank}}
\end{abstract}

\section{Introduction}
\label{sec:Introduction}

Information retrieval is a fundamental component within the field of natural language processing~\citep{chen-etal-2017-reading}. Retrieval aims to search a set of candidate documents from a large-scale corpus, and thus high recall retrieval with efficiency is required to cover more relevant documents as far as possible. Traditionally, retrieval has been dominated by lexical methods like TF-IDF and BM25~\citep{Robertson2009Probabilistic}, which treat queries and documents as sparse bag-of-words vectors and match them in token-level. Recently, neural networks have become prevalent to deal with information retrieval, where queries and documents are encoded into dense contextualized semantic vectors~\citep{Huang2020Embeddingbased, karpukhin-etal-2020-dense, ren-etal-2021-pair, Zhang2022ADVERSARIAL}, and then retrieval is performed with highly optimized vector search algorithms~\citep{Johnson2021BillionScale}.

Although numerous efforts have been dedicated to retrieval, the inherent efficiency requirement restricts the interaction between query and passage to a shallow level, leading to unsatisfactory retrieval results. Thus, in typical reranking~\citep{Nogueira2020Passage, sun-etal-2021-shot}, query and passage are concatenated and fed into a Transformer~\citep{Vaswani2017Attention} pre-trained on large corpus, to estimate a more fine-grained relevance score and further enhance the retrieval results with richer interaction. These methods consider each passage in isolation, ignoring the context of the retrieved passage list. Some learning to rank~\citep{Rahimi2016Building, William2008ListwiseApproach} and pseudo-relevance feedback~\citep{Zamani2016PseudoRelevance, Zhai2001ModelbasedFeedback} methods utilize the ordinal relationship or listwise context of retrieved documents to further refine the retrieval. Moreover, the necessity of integrating listwise context is confirmed in multi-stage recommendation systems~\citep{Liu2022Neurala}.

Inspired by the success of listwise modeling and collaborative filtering~\citep{Goldberg1992Using} in recommendation systems, we find that collaboration also exists among the passages in the retrieval list and has not been fully exploited. Intuitively, for a specific query, a set of passages relevant to the query tend to describe the same entities, events and relations~\citep{lee-etal-2019-latent}, while irrelevant ones outside of this set involve multifarious objects. Therefore, a passage is more likely to be relevant with the query if most of other passages share similar content with it. Similarities between passages can be naturally derived from retrievers, like BM25 scores in sparse\footnote{To stand out in contrast to dense retrieval, lexical retrieval is referred to as term sparse retrieval in this paper.} retrievers and dot product of embeddings in dense retrievers.

In addition, the sparse and dense retrieval methods emphasize distinct linguistic aspects. Sparse retrieval relies on lexical overlap while dense retrieval focuses on semantic and contextual relevance. Several researchers have attempted to integrate the merits of these two types of methods. \citet{karpukhin-etal-2020-dense}, \citet{Lin2020Distilling} and \citet{luan-etal-2021-sparse} exploit the linear combination of these two types of retrieval scores. \citet{seo-etal-2019-real}, \citet{Khattab2020ColBERT} and \citet{santhanam-etal-2022-colbertv2} index smaller units in sentence, \textit{i.e.}, words or phrases, to obtain fine-grained similarity. \citet{Gao2021Complementing} and \citet{yang-etal-2021-neural-retrieval} retrain dense retriever from scratch with the supervision of sparse signals. Nevertheless, the linear score combination lacks sufficient interaction, indexing smaller units sacrifices efficiency due to tremendous amount of embeddings, while rebuilding of retrievers discards their origin ranking capability.

To fully exploit the context of retrieved passages list and explore more sufficient ensemble of heterogeneous retriever, we propose a \textit{\textbf{Hyb}rid and Collaborative Passage Re\textbf{rank}ing} (HybRank) method, which leverages the collaboration within retrieved passages and incorporates diverse properties of retrievers for reranking.
Our method is a flexible plug-in reranker that can be applied to arbitrary passage lists, including those that have already been reranked by other methods.
In this work, without loss of generality, we employ the two most representative types of retrievers: sparse and dense retriever. Given a query and an initial retrieval list, we first extract similarities between them and a set of anchor texts via both the sparse and dense retrievers. We project and group them to form a set of hybrid and collaborative sequences, each corresponding to a query or passage. Afterwards, the relevance scores between the query and these passages are evaluated in the light of these sequences.

Extensive experiments demonstrate the consistent performance improvement brought by HybRank over passage lists from prevalent retrievers and strong rerankers. We elaborate ablation studies on the collaborative information, feature hybrid, anchor-wise interaction and the number of anchor passages, verifying the impact and indispensability of these components in HybRank.

\section{Method}

\begin{figure*}[t]
	\begin{center}
		\includegraphics[width=0.96\textwidth]{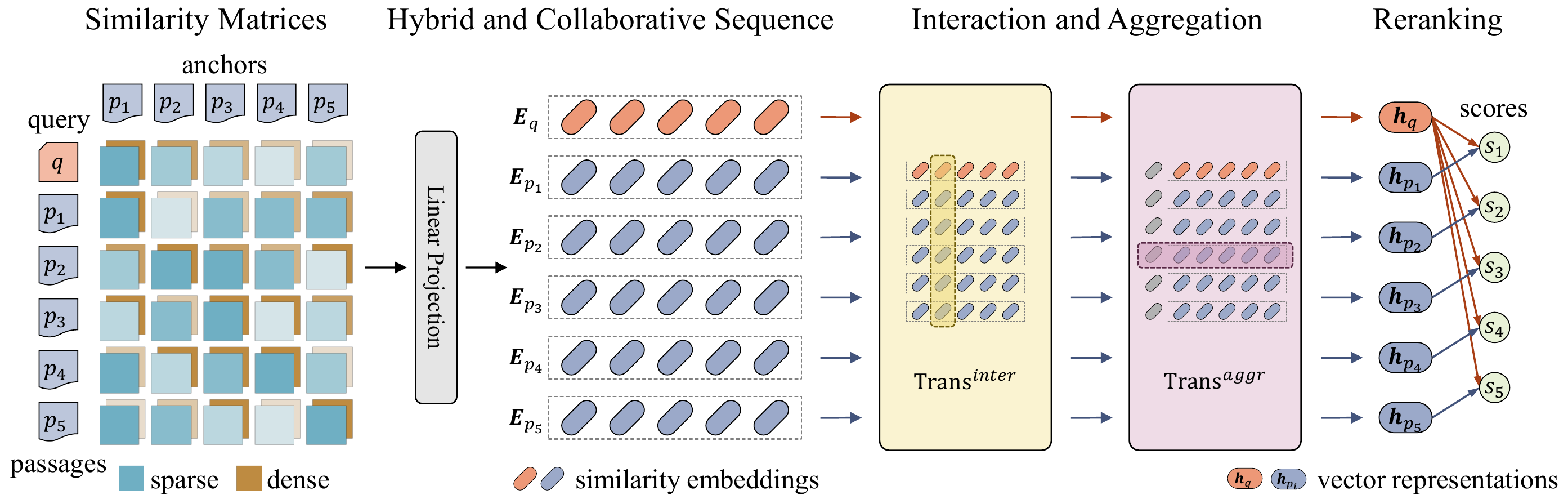}
	\end{center}
	\caption{Illustration of HybRank pipeline. For a specific query, the passage list is initialized by an arbitrary retriever. The passage list may have been reranked by another reranker before HybRank. We display a $5$-passage list as an example. First, similarities between query, passages and anchors are derived from sparse and dense retrievers. Then, these similarities are converted to hybrid and collaborative sequences as the representations of query and passages. Finally, these sequences are encoded into dense vectors via interaction and aggregation, and the reranking scores are obtained by dot product between the dense vectors of the query and each passage.}
	\label{fig:pipeline}
\end{figure*}

In mainstream information retrieval systems, the first-stage retrieval is designed to fetch a coarse candidate list from a large corpus $\mathcal{C}$. Inevitably, false positives, \textit{i.e.}, irrelevant passages in the retrieval list, are returned in the first-stage retrieval. To improve the precision of retrieval systems, the follow-up procedure reranking aims to distinguish the relevant passages from others in the retrieval list. This paper focuses on the reranking stage.

Formally, given a query $q$ and an initial passage list $\mathcal{P} = [p_1, p_2, \dots, p_N]$ from upstream retriever, the reranking task is to reorder the passage list by reassigning scores $\mathcal{S} = [s_1, s_2, \dots, s_N]$ for each of these passages. We denote positive passages in the list as $\mathcal{P}^+$ and negative ones as $\mathcal{P}^-$. In this section, we will present the details of HybRank. The pipeline of HybRank is illustrated in Figure~\ref{fig:pipeline}.

\subsection{Preliminaries}
\label{sec:prelimilaries}

\paragraph{Sparse Retrieval}
Traditionally, text retrieval is dominated by token-matching, where texts are encoded into high-dimensional sparse vectors using the statistic information of tokens. The most commonly-used sparse retrieval methods include TF-IDF, BM25 and so forth. We adopt BM25 score as the similarity metric of sparse retrieval due to its robustness and popularity.

Specifically, given the query $q$ and the document $d$, the BM25 score between $q$ and $d$ is obtained by summing the BM25 weights over the terms co-occurred in $q$ and $d$:
\begin{equation}
	\label{eqn:sparse}
	f^{s}(q, d) = \sum_{t \in q \cap d} w_t^{\mathrm{RSJ}} \frac{c_{t,d}}{k_1((1 - b) + b \frac{|d|}{l}) + c_{t,d}},
\end{equation}
where $t$ is a term, $w_t^{\mathrm{RSJ}}$ is $t$'s Robertson-Sp\"arck Jones weight, $c_{t,d}$ is the frequency of $t$ in $d$, $|d|$ is the document length and $l$ is the average length of all documents in the collection. $k_1$ and $b$ are tunable parameters. Refer to~\citet{Robertson2009Probabilistic} for more details about BM25.

\paragraph{Dense Retrieval}
Owning to the flexibility for a task-specific representation provided by learnable parameters, recent works leverage neural networks to encode text into dense vectors, and search similar documents for queries in vector space. Typically, the query and document are encoded separately, and the relevance score is measured by the similarity of their embeddings. Any neural architectures capable of encoding text into a single fixed-length vector are suitable for dense retrieval. We use the predominant Transformer~\citep{Vaswani2017Attention} encoder and dot product similarity, formulated as
\begin{equation}
	\label{eqn:dense}
	f^d(q, d) = \mathrm{T}_q(q)^\top \mathrm{T}_d(d),
\end{equation}
where $\mathrm{T}_q(\cdot)$ and $\mathrm{T}_d(\cdot)$ are Transformer encoders for queries and documents. Dot product similarity permits offline pre-encoding of large corpus and efficient retrieval via highly optimized vector nearest neighbor searching library~\citep{Johnson2021BillionScale}.

\subsection{Hybrid and Collaborative Sequence}
For a specific query, relevant passages tend to describe the same entities, events and relations from the query~\citep{lee-etal-2019-latent}. In other words, most passages in the retrieval list would resemble to the true positive ones. Inspired by the success of collaborative filtering~\citep{Goldberg1992Using} in recommendation systems, we utilize the similarities between passages to distinguish the positive passages in the retrieval list.

\paragraph{Collaborative Sequence}
Similarity measurements can be naturally derived from retrievers. \textit{e.g.}, BM25 score in sparse retriever and dot product in dense retriever as described in Section~\ref{sec:prelimilaries}. We compute the similarity between each passage and a set of anchors, which are the top-$L$ passages of the retrieval list in this work and will collaborate to distinguish the positive passages. These similarity scores between passages can be pre-computed, as HybRank utilizes off-the-shelf retrievers. Denoting similarity score between passage $p_i$ and $p_j$ as $f_{ij} \in \R$, the passage $p_i$ can be represented as a sequence of similarity scalars $\vx_{p_i} = [f_{i1}, f_{i2}, \dots, f_{iL}] \in \R^L$. 

Nevertheless, according to our observation, the similarity scalars within a retrieval list tend to concentrate on a small range. This is a reasonable phenomenon for that retrievers fetch relatively similar passages from the large corpus. To obtain more distinctive features, we employ a temperature $\softmax$ to stretch the distribution of similarities. After that, a $\min$-$\max$ normalization is applied to scale them into range $[-1, 1]$. These two transforms are formulated as
\begin{equation}
	\label{eqn:norm}
	\begin{split}
		\vx &= \softmax(\vx / t), \\
		\vx &= 2 \cdot \frac{\vx - \min(\vx)}{\max(\vx) - \min(\vx)} - 1,
	\end{split}
\end{equation}
where $t$ is the temperature. Subscripts of $\vx_{p_i}$ are omitted for brevity.

\paragraph{Feature Hybrid}
Similarity metrics of sparse and dense retrievers concentrate on lexical overlap and semantic relevance, respectively. To combine the lexical and semantic properties embedded in sparse and dense retrievers, we mix their similarity scores\footnote{In this paper, we refer to similarity score from sparse and dense retrievers as sparse similarity and dense similarity, respectively.} by stacking them in a channel manner. Formally, we substitute the similarity scalar $f_{ij}$ in $\vx_{p_i}$ with a vector $\vx_{ij} = [f_{ij}^s, f_{ij}^d] \in \R^2$, where $f_{ij}^s$ is the sparse similarity computed as Eqn.~\ref{eqn:sparse} and $f_{ij}^d$ is the dense similarity computed as Eqn.~\ref{eqn:dense}. After that, the representation of passage $p_i$ is turned into a sequence of similarity vectors $\mX_{p_i} = [\vx_{i1}, \vx_{i2}, \dots, \vx_{iL}] \in \R^{L \times 2}$. Additionally, we map these similarity vectors in the sequence to $D$ dimension with a trainable linear projection:
\begin{equation}
	\ve_{ij} = \vx_{ij} \mW,
\end{equation}
where $\mW \in \R^{2 \times D}$ is a learnable parameter and $\ve_{ij} \in \R^{D}$ are embedded similarities. Thereafter, passage $p_i$'s representation becomes a sequence of similarity embeddings $\mE_{p_i} = [\ve_{i1}, \ve_{i2}, \dots, \ve_{iL}] \in \R^{L \times D}$, which comprises the similarity information between $p_i$ and anchor passages originating from both sparse and dense retrievers. These similarities deliver substantial information for the collaboration of passages and hold both the lexical and semantic properties from retrievers. With the same procedure, we compute the similarities between query and anchors, and derive the query representation $\mE_{q} = [\ve_{q1}, \ve_{q2}, \dots, \ve_{qL}] \in \R^{L \times D}$. Noted that the similarities from sparse and dense retriever are stretched and normalized individually before linear projection, as described in Eqn.~\ref{eqn:norm}.

Consequently, we obtain $N + 1$ collaborative sequences in total, each representing a passage or a query and consisting of their lexical and semantic similarity information with $L$ anchor passages.

\subsection{Interaction and Aggregation}
Following the prevalent sequence similarity learning paradigm in the field of natural language processing~\citep{reimers-gurevych-2019-sentence, gao-etal-2021-simcse}, we expect to measure the relevance of query and passage with their collaborative sequences in vector space. We obtain these vector representations by anchor-wise interaction and sequence aggregation in HybRank.
\paragraph{Anchor-wise Interaction}
The $j$-th elements $\ve_{*j}$ in these collaborative sequences $\mE_*$ indicate the similarities between retrieved passages and the $j$-th anchor passage. The importance of these anchors varies since they are picked with a single strategy. Specifically, an anchor is worthy of more consideration if showing strong correlation with a majority of retrieved passages, and vice versa.

To assess the quality of anchor passages, we conduct anchor-wise interaction. Concretely, for each position $j$, we collect the $j$-th similarity embedding $\ve_{*j}$ from query sequence and every passage sequences and refine them with a Transformer encoder, denoted as
\begin{equation}
	\begin{aligned}
	\ve'_{qj}, &\ve'_{1j}, \ve'_{2j}, \dots, \ve'_{Nj} \\
	& = \mathrm{Trans}^{inter}(\ve_{qj}; \ve_{1j}; \ve_{2j}; \dots; \ve_{Nj}),
	\end{aligned}
\end{equation}
where $\ve'_{*j} \in \R^{D}$. Position embeddings are added to $\ve_{*j}$ according to its rank ``$*$'' for retaining the passage rank information. Subsequently, the similarity embedding sequences $\mE_{*}$ are converted to $\mE'_{*} = [\ve'_{*1}, \ve'_{*2}, \dots, \ve'_{*L}]$ and enhanced with the importance information of anchor passages.

\paragraph{Sequence Aggregation}
We encode these sequences into dense vectors by aggregating the enhanced similarity embeddings. To be specific, we prepend a $\mathrm{[CLS]}$ embedding to the collaborative sequence, feed the extended sequence into another Transformer encoder and use the output of $\mathrm{[CLS]}$ as the representation of $p_i$, formulated as
\begin{equation}
	\vh_{p_i} = \mathrm{Trans}^{aggr}(\mathrm{[CLS]} \oplus \mE'_{p_i})_{\mathrm{[CLS]}},
\end{equation}
where $\mathrm{[CLS]} \in \R^{1 \times D}$, $\mE'_{p_i} \in \R^{L \times D}$ and $\oplus$ denotes the concatenation operation. $\vh_{p_i} \in \R^{D}$ is the vector representation of passage $p_i$. The query representation $\vh_q \in \R^{D}$ is derived analogously.

\paragraph{Receptive Field and Complexity}
Interestingly, from another perspective, the anchor-wise interaction and sequence aggregation are equivalent to a column-wise and a row-wise attention applied on the matrix formulated by similarities of query, passages and anchors. Global receptive field is provided by these two axial-wise attentions~\citep{Ho2019Axial}. Consequently, similarity vector $\vx_{ij}$ perceives with each other, and the vector representations of query and passages are aware of the collaborative information from others.

A more direct approach to obtain global receptive field is element-wise interaction. Concretely, we can feed the concatenation of all sequences $\mE$ into a single Transformer encoder, and output representations for each passage and query via multiple separate $\mathrm{[CLS]}$ tokens. However, due to the self-attention operation in Transformer, the computational complexity of element-wise interaction achieves $O(N^2 L^2)$. In contrast, our method reduce the complexity to $O(N^2 L + N L^2)$, by decomposing the element-wise attention on the similarity matrix into axial-wise. Note that the complexity can be further reduced to $O(N L + N L)$ if leveraging linear Transformers~\citep{pmlr-v119-katharopoulos20a, linformer} instead of vanilla Transformers.

\subsection{Reranking and Training}
\paragraph{Reranking} 
Considering that query and passages have been converted into dense vectors encoded with collaborative information, we have several alternatives to judge the vector similarity as the relevance score between the query and passage. We use dot product in this work and thus the relevance score between query $q$ and $p_i$ is computed by
\begin{equation}
	s_i = \vh_q^\top \vh_{p_i}.
\end{equation}
Then passages are sorted in descending order of their relevance score $s_i$ with query.

\paragraph{Training}
In order to assign high scores to relevant passages and low scores to irrelevant ones, HybRank needs to pull together the representation of relevant passages and query, while push the representation of irrelevant ones as apart from the query as possible. As there may exist more than one positive passage in the list, vanilla softmax loss fails to be directly applied to HybRank. We adopt the supervised contrastive loss~\citep{Khosla2020Supervised} to cope with multiple positives, which performs summation over positives outside the log function in softmax. The loss is formulated as
\begin{equation}
	\mathcal{L}(q, \mathcal{P}) = -\frac{1}{|\mathcal{P}^+|} \sum_{p_i \in \mathcal{P}^+} \log \frac{\exp (s_i / \tau)}{\sum_{p_j \in \mathcal{P}} \exp (s_j / \tau)},
\end{equation}
where $|\mathcal{P}^+|$ is the number of positive passages in the retrieval list, and $\tau$ is a tunable temperature.

\section{Experiments}

\subsection{Datasets}

\paragraph{Natural Questions}
\citep{kwiatkowski-etal-2019-natural} consists of real English questions from Google search engine with golden passages from English Wikipedia pages and answer span annotations. Following the settings from~\citet{karpukhin-etal-2020-dense}, we report the test set top-$k$ accuracy (R@k), which evaluates the percentage of queries whose top-$k$ retrieved passages contain the answers.

\paragraph{MS MARCO}
\citep{Bajaj2018MS} includes English queries from Bing search logs and was originally designed for machine reading comprehension. Following previous works~\citep{qu-etal-2021-rocketqa, ren-etal-2021-rocketqav2}, we evaluate the dev set R@k as well as Mean Reciprocal Rank (MRR), which means the average reciprocal of the first retrieved relevant passage rank.

\paragraph{TREC 2019/2020}
\citep{TREC2019,TREC2020} originate from TREC 2019/2020 Deep Learning (DL) Track. These two tracks provide additional Bing search queries and require to retrieve passages from the MS MARCO corpus. We use the official setting and evaluate the NDCG@10 of HybRank trained on MS MARCO with their test set.

\subsection{Implementation Details}
HybRank is a flexible plug-in reranker which can be applied on arbitrary passage lists including those that have already been reranked by other methods. Thus, we test HybRank against not only retrieval systems but also systems with other rerankers in it. We adopt dense retrievers which outperform sparse ones after elaborated pre-training~\citep{Chang2020PRETRAINING, gao-callan-2021-condenser, gao-callan-2022-unsupervised} and fine-tuning~\citep{sachan-etal-2021-end}, as well as strong cross-encoder based rerankers, to initialize the passage list. We simply select all passages in the initial list as anchors. The impact of anchor passages will be discussed in Section~\ref{sec:analysis}. These methods are implemented using RocketQA toolkit\footnote{\url{https://github.com/PaddlePaddle/RocketQA}.} and Pyserini toolkit~\citep{Lin2021Pyserini} which is built on Lucene\footnote{\url{https://lucene.apache.org}.} and FAISS~\citep{Johnson2021BillionScale}.

The hyper-parameters in HybRank are as follows. The temperature $t$ in the feature normalization is set to $100$ and $10$ for sparse and dense similarity, respectively. We randomly initialize a $2$-layer Transformer encoder for $\mathrm{Trans}^{inter}$ and $1$-layer for $\mathrm{Trans}^{aggr}$ using Huggingface Transformers~\citep{wolf-etal-2020-transformers}. The embedding dimension, MLP inner-layer dimension and number of heads are $64$, $256$ and $8$, respectively. There are 0.22M parameters in total. The temperature $\tau$ in the loss function is $0.07$. We adopt the Adam optimizer with an initial learning rate $1 \times {10}^{-3}$ with the warm-up ratio $0.1$, followed by a cosine learning rate decay. We use gradient clipping of $2$ and weight decay of ${1 \times 10}^{-6}$.  We train the model for $100$ epochs with batch size $32$, which takes about $13$ hours on Natural Questions and $4$ days on MS MARCO. All experiments are conducted on a single NVIDIA RTX 3090 GPU.

\subsection{Results}
\label{sec:results}

\begin{table*}[t]
	\centering
	\small
	\begin{tabular}{llll}
		\toprule
		& \multicolumn{3}{c}{\textbf{Natural Questions Test}} \\
		\cmidrule(lr){2-4}
		& \multicolumn{1}{c}{R@1} & \multicolumn{1}{c}{R@5} & \multicolumn{1}{c}{R@20} \\
		\midrule
		DPR-Multi + HybRank & 45.82 $\rightarrow$ 51.99 (\textbf{+6.17}) & 68.12 $\rightarrow$ 72.71 (\textbf{+4.59}) & 80.31 $\rightarrow$ 83.24 (\textbf{+2.93}) \\
		\midrule
		DPR-Single + HybRank & 47.95 $\rightarrow$ 53.13 (\textbf{+5.18}) & 69.39 $\rightarrow$ 73.05 (\textbf{+3.66}) & 80.97 $\rightarrow$ 82.99 (\textbf{+2.02}) \\
		\midrule
		FiD-KD + HybRank & 50.36 $\rightarrow$ 52.85 (\textbf{+2.49}) & 74.10 $\rightarrow$ 74.46 (\textbf{+0.36}) & 84.27 $\rightarrow$ 84.49 (\textbf{+0.22}) \\
		\midrule
		ANCE + HybRank & 52.66 $\rightarrow$ 53.63 (\textbf{+0.97}) & 72.66 $\rightarrow$ 73.57 (\textbf{+0.91}) & 83.05 $\rightarrow$ 83.88 (\textbf{+0.83}) \\
		\midrule
		RocketQA-retriever + HybRank & 51.74 $\rightarrow$ 56.07 (\textbf{+4.33}) & 74.02 $\rightarrow$ 77.04 (\textbf{+3.02}) & 83.99 $\rightarrow$ 85.68 (\textbf{+1.69}) \\
		\midrule
		RocketQA-reranker + HybRank & 54.60 $\rightarrow$ 59.83 (\textbf{+5.23}) & 76.59 $\rightarrow$ 78.73 (\textbf{+2.14}) & 85.01 $\rightarrow$ 86.40 (\textbf{+1.39}) \\
		\midrule
		RocketQAv2-retriever + HybRank & 55.57 $\rightarrow$ 56.98 (\textbf{+1.41}) & 75.98 $\rightarrow$ 76.65 (\textbf{+0.67}) & 84.46 $\rightarrow$ 85.76 (\textbf{+1.30}) \\
		\midrule
		RocketQAv2-reranker + HybRank & 57.17 $\rightarrow$ 59.50 (\textbf{+2.33}) & 75.98 $\rightarrow$ 78.34 (\textbf{+2.36}) & 84.71 $\rightarrow$ 86.26 (\textbf{+1.55}) \\
		\bottomrule
	\end{tabular}
	\caption{The reranking performance of HybRank on Natural Questions from a single run. We build HybRank upon DPR~\citep{karpukhin-etal-2020-dense}, FiD-KD~\citep{Izacard2021DISTILLING}, ANCE~\citep{Xiong2021APPROXIMATE}, RocketQA~\citep{qu-etal-2021-rocketqa} and RocketQAv2~\citep{ren-etal-2021-rocketqav2}. The performance of these baselines and HybRank are on the left and right side of arrows, respectively. Improvements brought by HybRank are highlighted in bold.}
	\label{tab:main_results_NQ}
\end{table*}

\begin{table*}[t]
	\centering
	\small
	\begin{tabular}{llll}
		\toprule
		& \multicolumn{1}{c}{\textbf{MS MARCO Dev}} & \multicolumn{1}{c}{\textbf{TREC 2019}} & \multicolumn{1}{c}{\textbf{TREC 2020}} \\
		\cmidrule(lr){2-2} \cmidrule(lr){3-3} \cmidrule(lr){4-4}
		\multirow{2}{*}{} & \multicolumn{1}{c}{MRR@10} & \multicolumn{1}{c}{NDCG@10} & \multicolumn{1}{c}{NDCG@10} \\
		\midrule
		DistilBERT-KD + HybRank & 32.50 $\rightarrow$ 36.24 (\textbf{+3.74}) & 69.23 $\rightarrow$ 72.55 (\textbf{+3.32}) & 60.58 $\rightarrow$ 66.71 (\textbf{+6.13}) \\
		\midrule
		ANCE + HybRank & 33.01 $\rightarrow$ 36.44 (\textbf{+3.43}) & 62.37 $\rightarrow$ 70.41 (\textbf{+8.04}) & 60.00 $\rightarrow$ 63.70 (\textbf{+3.70}) \\
		\midrule
		TCT-ColBERT-v1 + HybRank & 33.49 $\rightarrow$ 36.23 (\textbf{+2.74}) & 65.42 $\rightarrow$ 73.21 (\textbf{+7.79}) & 61.03 $\rightarrow$ 66.91 (\textbf{+5.88}) \\
		\midrule
		TAS-B + HybRank & 34.44 $\rightarrow$ 36.38 (\textbf{+1.94}) & 70.49 $\rightarrow$ 74.82 (\textbf{+4.33}) & 63.89 $\rightarrow$ 66.53 (\textbf{+2.64}) \\
		\midrule
		TCT-ColBERT-v2 + HybRank & 35.85 $\rightarrow$ 37.55 (\textbf{+1.70}) & 71.15 $\rightarrow$ 74.06 (\textbf{+2.91}) & 64.32 $\rightarrow$ 66.35 (\textbf{+2.03}) \\
		\midrule
		RocketQA-retriever + HybRank & 35.77 $\rightarrow$ 36.97 (\textbf{+1.20}) & 70.49 $\rightarrow$ 74.79 (\textbf{+4.30}) & 63.74 $\rightarrow$ 67.25 (\textbf{+3.51}) \\
		\midrule
		RocketQA-reranker + HybRank & 40.51 $\rightarrow$ 40.98 (\textbf{+0.47}) & 75.40 $\rightarrow$ 77.05 (\textbf{+1.65}) & 67.66 $\rightarrow$ 69.85 (\textbf{+2.19}) \\
		\midrule
		RocketQAv2-retriever + HybRank & 37.28 $\rightarrow$ 38.74 (\textbf{+1.46}) & 70.14 $\rightarrow$ 73.63 (\textbf{+3.49}) & 63.04 $\rightarrow$ 67.87 (\textbf{+4.83}) \\
		\midrule
		RocketQAv2-reranker + HybRank & 41.15 $\rightarrow$ 41.40 (\textbf{+0.25}) & 73.24 $\rightarrow$ 74.92 (\textbf{+1.68}) & 69.47 $\rightarrow$ 70.71 (\textbf{+1.24}) \\
		\bottomrule
	\end{tabular}
	\caption{The reranking performance of HybRank on MS MARCO and TREC 2019/2020 from a single run. We built HybRank upon DistilBERT-KD~\citep{Hofstatter2021Improving}, ANCE~\citep{Xiong2021APPROXIMATE}, TCT-ColBERT-v1~\citep{Lin2020Distilling}, TAS-B~\citep{Hofstatter2021Efficiently}, TCT-ColBERT-v2~\citep{lin-etal-2021-batch}, RocketQA~\citep{qu-etal-2021-rocketqa} and RocketQAv2~\citep{ren-etal-2021-rocketqav2}. The performance of these baselines and HybRank are on the left and right side of arrows, respectively. Improvements brought by HybRank are highlighted in bold.}
	\label{tab:main_results_MSMARCO}
\end{table*}

Table~\ref{tab:main_results_NQ} and Table~\ref{tab:main_results_MSMARCO} summarize the performance of HybRank and baselines on the Natural Questions, MS MARCO and TREC 2019/2020 datasets. More detailed evaluation results are listed in Appendix~\ref{appdx:full_results}. Some of adopted retrieval baselines involve both sparse and dense similarity from different perspectives. 
DPR~\citep{karpukhin-etal-2020-dense} selects hard negative samples from passages returned by BM25; FiD-KD~\citep{Izacard2021DISTILLING} starts its iterative training with passages retrieved using BM25; TCT-ColBERT-v1~\citep{Lin2020Distilling} proposes an alternative approximation for linear combination of dense and sparse retrieval; TCT-ColBERT-v2~\citep{lin-etal-2021-batch} further studies the dense-sparse hybrid in terms of quality, time and space. Besides, ANCE~\citep{Xiong2021APPROXIMATE} discovers new negatives via nearest neighbor search during model training; TAS-B~\citep{Hofstatter2021Efficiently} proposes balanced sampling strategies to compose informative training batches; DistilBERT-KD~\citep{Hofstatter2021Improving} leverages cross-architecture knowledge distillation for model-agnostic training. 

From the results we can observe that HybRank shows a consistent improvements over upstream retrievers and even rerankers. In general, HybRank based on stronger baselines can produce better reranking results. For example, HybRank built upon the retriever of RocketQA outperforms HybRank built upon DPR-Multi on Natural Questions, and the same phenomenon can be observed on most retrievers. Additionally, HybRank built upon systems with reranker further improves the performance on both datasets. These results prove the advantage of reranking based on arbitrary off-the-shelf retrievers and even other reranked results, which distinguishes HybRank from other rerankers.

The most surprising aspect of these results is that, in spite of inferior reranking results, low-scoring retrievers gain more relative improvements from HybRank than high-scoring ones. This result may be explained by the fact that HybRank relies heavily on the complementary information provide by sparse similarity. Low-scoring retrievers receive relatively more valuable information from sparse similarity than high-scoring retrievers, and accordingly improve more performance over upstream retrievers. We will discuss more on sparse-dense hybrid in Section~\ref{sec:analysis}.

\subsection{Analysis}
\label{sec:analysis}

In this section, we discuss the impact of core components of HybRank: the hybrid and collaborative features, the anchor-wise interaction and the number of anchor passages. All experiments are conducted on Natural Questions dataset with DPR-Multi retriever.

\paragraph{Collaborative Feature}

\begin{table}[t]
	\centering
	\small
	\begin{tabular}{cccccc}
		\toprule
		& \multicolumn{1}{c}{R@1} & \multicolumn{1}{c}{R@5} & \multicolumn{1}{c}{R@10} & \multicolumn{1}{c}{R@20} & \multicolumn{1}{c}{R@50} \\
		\midrule
		retriever & 45.82 & 68.12 & 75.24 & 80.30 & 84.57 \\
		r/d anchor & 46.18 & 68.84 & 75.43 & 80.91 & 85.01 \\
		w/o $q$-$p$ & 47.12 & 69.17 & 75.54 & 80.47 & 85.07 \\
		w/o inter & 49.92 & 69.61 & 76.32 & 81.02 & 84.99 \\
		w/o collab & 50.78 & \textbf{72.91} & \textbf{79.28} & 83.10 & 85.79 \\
		HybRank & \textbf{51.99} & 72.71 & 79.03 & \textbf{83.24} & \textbf{85.93} \\
		\bottomrule
	\end{tabular}
	\caption{The results of ablation study for collaborative features, anchor-wise interaction and anchor passages on the test set of Natural Questions.}
	\label{tab:ablation_collaborative}
\end{table}

The main difference between HybRank and other works is, it leverages the collaborative information between retrieved passages. To verify the impact of passage collaboration on reranking, we omit the collaborative feature in ``w/o collab'' by substituting only query-passage similarities for collaborative sequences, \textit{i.e.}, representing each passage as a one-token sequence according to its similarities with query. Besides, we exclude the query-passage similarity in ``w/o $q$-$p$'' by representing query via a learnable token rather than aggregated collaborative sequence. The results are presented in Table~\ref{tab:ablation_collaborative}, where ``retriever'' denotes the assessment of initial passage list.

Table~\ref{tab:ablation_collaborative} indicates that ``w/o collab'' shows an appreciable gain over ``retriever'', demonstrating that query-passage similarity is an essential and indispensable feature for HybRank. The most remarkable phenomenon is, ``w/o $q$-$p$'' surpasses ``retriever'' by a large margin, despite the fact that ``w/o $q$-$p$'' is completely unaware of the query. Namely, HybRank has the ability to distinguish the positive even only with the collaborative information among passages. Furthermore, standing on the shoulder of query-passage similarity, HybRank achieves even better results than ``w/o collab'', which sufficiently substantiates the reranking capability of collaborative information.

\paragraph{Anchor-wise Interaction}
Apart from the collaborative sequence, anchor-wise interaction provides extra collaboration between sequences. We eliminate the $\mathrm{Trans}^{inter}$ and directly aggregate the linear projected collaborative sequence to study the effectiveness of anchor-wise interaction.

Table~\ref{tab:ablation_collaborative} shows that there is a noticeable drop of performance without anchor-wise interaction. The discrepancy could be attributed to the restricted receptive field. ``w/o inter'' individually encodes each collaborative sequences of query and passages into dense vectors without anchor-wise interaction. In this manner, the relevance of these sequences is evaluated only in vector space where sequence information are severely compressed and not expressive enough. In contrast, equipped with anchor-wise interaction, HybRank is capable of obtaining a global receptive field. Each element in these sequences captures the context of elements in all sequences, enabling more informative vector representation and fine-grained relevance estimation.

\paragraph{Feature Hybrid}
\begin{table}[t]
	\centering
	\small
	\setlength\tabcolsep{4pt}
	\begin{tabular}{ccccccc}
		\toprule
		list & feature & \multicolumn{1}{c}{R@1} & \multicolumn{1}{c}{R@5} & \multicolumn{1}{c}{R@10} & \multicolumn{1}{c}{R@20} & \multicolumn{1}{c}{R@50} \\
		\midrule
		\multirow{4}*{sparse} & none & 23.82 & 45.18 & 55.54 & 63.93 & 73.55 \\
		& sparse & 30.50 & 50.39 & 59.00 & 67.26 & 75.24 \\
		& dense & 47.01 & 64.68 & \textbf{70.39} & \textbf{74.49} & \textbf{77.81} \\
		& hybrid & \textbf{47.15} & \textbf{64.82} & 69.78 & 74.32 & 77.65 \\
		\midrule
		\multirow{4}*{dense} & none & 45.82 & 68.12 & 75.24 & 80.30 & 84.57 \\
		& dense & 46.70 & 68.45 & 75.04 & 80.19 & 84.88 \\
		& sparse & 50.89 & 71.86 & 78.98 & 83.16 & 85.90 \\
		& hybrid & \textbf{51.99} & \textbf{72.71} & \textbf{79.03} & \textbf{83.24} & \textbf{85.93} \\
		\bottomrule
	\end{tabular}
	\caption{The results of ablation study for feature hybrid on the test set of Natural Questions.}
	\label{tab:ablation_hybrid}
\end{table}

Despite the fact that the similarities of sparse and dense retriever reflect different aspect of linguistics, \textit{i.e.}, lexical overlap and semantic relevance, both of them tend to have collaborative property. Hence, it is more natural and easier to mix sparse and dense retrieval from the perspective of collaboration. To illustrate the complementarity of sparse and dense features and the necessity of feature hybrid in HybRank, we separately validate the effect of the two individual features and their hybrid. The ablations are conducted not only on initial passage list retrieved by dense retriever, but also list retrieved by sparse retriever for integrity and comparison.

Identical trends can be observed from two settings of experiments in Table~\ref{tab:ablation_hybrid}. The performance gains are limited when retrievers used for passage retrieval and similarity computation are same, but dramatically increase when they are different. Furthermore, additional slight improvements can be seen with the hybrid of the two features on both settings. These phenomena reveal that the main performance gains originate from the retriever different with that in retrieval stage, while the same type only plays an auxiliary role. Consequently, we draw the credible conclusion that different types of similarities provide additional complementary information over the initial passage list. 

Moreover, regardless of feature used, HybRank achieves better results on passage list retrieved by dense retriever than sparse one, as more positives are contained in the dense retrieved list. This also corroborates the findings of Section~\ref{sec:results} that superior initial passage list leads to better reranking results with HybRank.

\paragraph{Number of Anchor Passages}

\begin{figure*}[t]
	\centering
	\begin{minipage}[t]{0.96\textwidth}
		\centering
		\includegraphics[width=1\textwidth]{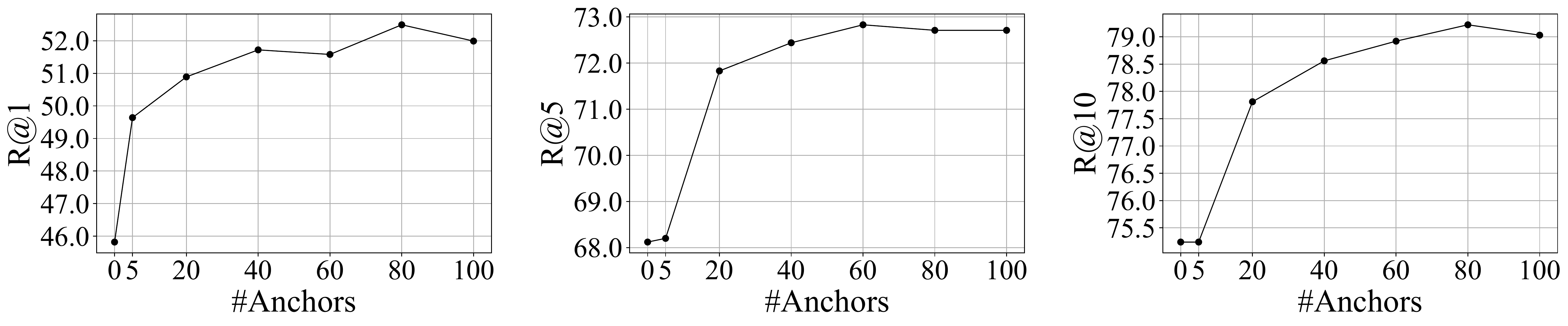}
	\end{minipage}
	\begin{minipage}[t]{0.64\textwidth}
		\centering
		\includegraphics[width=1\textwidth]{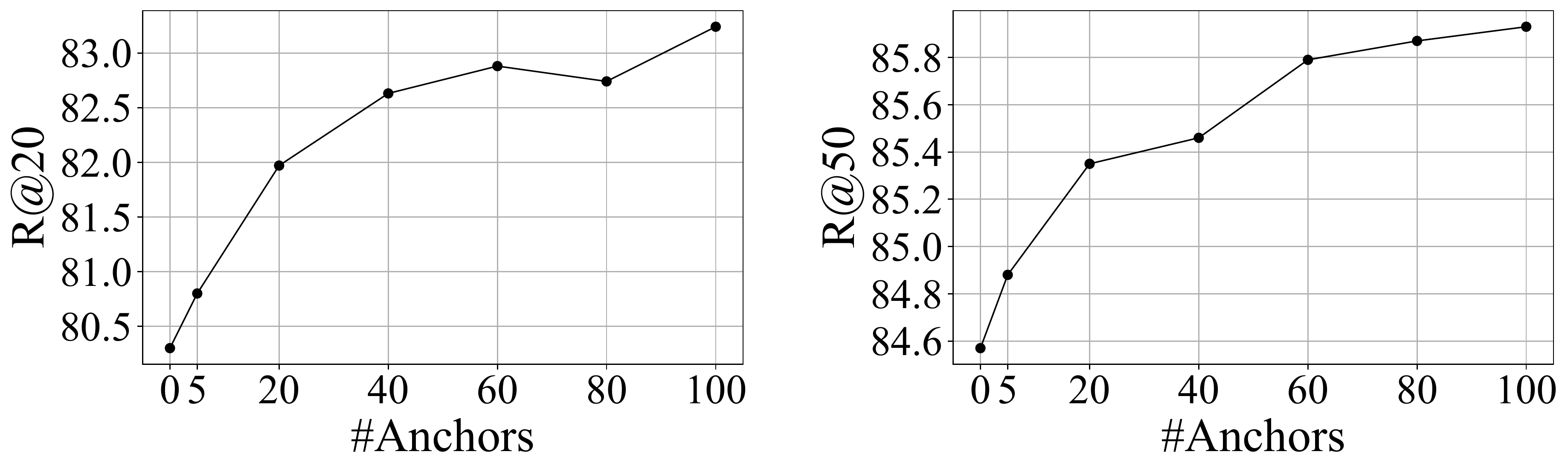}
	\end{minipage}
	\caption{Impact study on the number of anchor passages. We conduct experiments on the test set of Natural Questions with anchor number $5, 20, 40, 60, 80, 100$. The metric of anchor number $0$ denotes the assessment of initial retrieval list.}
	\label{fig:ablation_anchors}
\end{figure*}

We evaluate the performance of HybRank under different number of anchors to study its impact. What can be clearly seen in Figure~\ref{fig:ablation_anchors} is a consistent growth of performance as the anchor number $L$ increasing. The underlying philosophy is that, with more anchor passages the passage list can derive more agreement to facilitate the collaboration between passages and alleviate the distraction from noisy ones. The positive correlation between the performance and anchor number indicates the effect of collaborative information in the retrieval list. 

Despite the consistent growth with anchor number, the rate of performance increase begins to slow down when the number of anchors is greater than $60$. Anchor passages are used for deriving collaborative information, and thus with more diverse anchors we can obtain more distinctive collaborative features. As the anchor number approaches to $100$, the diversity of passages levels off, leading to stable performance with larger anchor numbers.

As $L$ increase to a very large number, the average relevance of anchors will degrade to a low level. A legitimate concern may be that poor quality anchor set would pollute the collaborative aspect. Due to the $O(L^2)$ computational complexity of sequence aggregation in HybRank, it is hard to directly perform experiments on large $L$. But we simulate the poor quality anchor set by randomly selecting anchor passages from corpus $\mathcal{C}$. ``r/d anchor'' in Table~\ref{tab:ablation_collaborative} indicates that random anchors slightly improves the performance but still lags far behind the relevant anchors, demonstrating the benefits of collaborative information and the predominance of the anchor quality.

Nevertheless, the selection of anchor passages is flexible. Ideally, more elaborated anchor passage selection, \textit{e.g.}, clustering the passages from the corpus and selecting a fixed number of clustering centroids as anchors, would further enhance the performance and efficiency of HybRank. We leave the exploration of other anchor selecting strategy as a future work.

\section{Related Work}

\subsection{Text Retrieval}
Retrieval is the first stage of information retrieval which requires high recall to cover more relevant document in the retrieval list. Traditional sparse approaches like TF-IDF and BM25~\citep{Robertson2009Probabilistic} rely on lexical overlap between query and documents.
Although having dominated the field of text retrieval for a long time, these sparse methods suffer from lexical gap~\citep{Berger2000Bridging}, namely, the synonymy problem. 
To tackle this issue, earlier techniques~\citep{Nogueira2019Document, Dai2020ContextAware} adopt neural networks to reinforce the sparse methods. Recently proposed dense retrieval approaches~\citep{karpukhin-etal-2020-dense, Xiong2021APPROXIMATE} directly encode the query and passages into dense vectors via dual-encoder, which capture semantic in text and enable low-latency search via highly optimized algorithms, \textit{e.g.}, FAISS~\citep{Johnson2021BillionScale}.

These two types of methods are not mutually exclusive and one's weakness is the other's strength.
Some researchers combine the sparse and dense methods by score ensemble, improved training or trade-off model between sparse and dense retriever.
\citet{karpukhin-etal-2020-dense} samples hard negatives from sparse retriever for the training of dense retriever. \citet{seo-etal-2019-real}, \citet{Khattab2020ColBERT} and \citet{santhanam-etal-2022-colbertv2} index terms or phrases instead of documents for more fine-grained similarity and higher efficiency.
\citet{Lin2020Distilling} and \citet{luan-etal-2021-sparse} explore the linear sparse-dense score combination and its alternatives.
\citet{Gao2021Complementing} and \citet{yang-etal-2021-neural-retrieval} leverages the lexical matching or token-level interaction signals to train the dense retriever.

However, among these methods, score ensemble lacks sufficient interaction of sparse and dense methods, smaller units indexing sacrifices efficiency, and retraining one type of retriever with the help of the other discards its origin ranking capability. In contrast, our method can be applied to arbitrary passage list, incorporating the lexical and semantic properties of off-the-shelf retrievers and meanwhile ensuring the generality and flexibility.

\subsection{Text Reranking}
The second stage reranking is based on the results of retrieval system and aims to create a more fine-grained comparison within retrieval list. Typically, cross-encoder is utilized to capture the interactions between query and passage in token-level. \citet{Nogueira2020Passage} and \citet{sun-etal-2021-shot} adopt BERT~\citep{devlin-etal-2019-bert} to achieve token-level interactions with attention mechanism~\citep{Vaswani2017Attention}. To reduce the massive computation overhead~\citep{reimers-gurevych-2019-sentence}, \citet{Khattab2020ColBERT} and \citet{gao-etal-2020-modularized} propose a lightweight interaction on dense representations from retrievers. While based on first-stage retrieval, these methods individually compute the relevance for each retrieved passage, omitting the extra information implied by the whole list and requiring multiple runs. 

Several pseudo-relevance feedback approaches~\citep{He2009FindingGood, Zamani2016EstimatingEmbedding, Zamani2016PseudoRelevance} aim to refine the query model with the top-retrieved documents. Listwise context is also well explored in multi-stage recommendation systems~\citep{Liu2022Neurala}, such as PRM~\citep{Pei2019PersonalizedRe}, which regards each item as a token, learns the mutual influence between items using self-attention and reranks all items altogether. Different from prior approaches, we extract the collaborative feature from the retrieval list, represent the query and each passages as hybrid and collaborative sequences, and measure the relevance between query and passages using these sequences from the perspective of collaboration.

\section{Conclusion}
We introduce HybRank, a hybrid and collaborative passage reranking method. HybRank extracts the similarities between texts via off-the-shelf retrievers to form hybrid and collaborative sequences as the representations of query and passages. Efficient reranking is based on these sequences which incorporate the lexical and semantic properties of sparse and dense retrievers. Extensive experiments confirm the effectiveness of HybRank upon arbitrary passage list. Elaborated ablation studies investigate the impact of core components in HybRank. We hope our work could provide inspiration for researchers in the field of information retrieval, and steer more exploration on collaboration and correlation between texts.

\section*{Limitations}
We evaluate HybRank on Natural Questions, MS MARCO and TREC 2019/2020 datasets, which focus on English Open-domain Question Answering. Although none of the components in HybRank are specifically designed for English, the verification of HybRank on other languages is limited. Otherwise, there are more general information retrieval tasks involving diversity or broader coverage in the returned results. Considering the possibility of lacking collaborative property, whether HybRank can generalize to these high-coverage retrieval tasks is still inconclusive.

As Transformer encoder architecture is adopted in the sequence interaction and aggregation, the computation cost would be unacceptable when the length of passage list or number of anchors is too large. This is also the reason why we only conduct experiments with anchor numbers no more than $100$. Besides, HybRank only uses similarities computed by off-the-shelf retrievers as input features, and thus lacks sufficient interaction between raw inputs. The performance of HybRank may be limited by the capability of upstream retrievers. How to incorporate the interaction of raw inputs into HybRank while avoiding massive computation cost is still an open problem for further investigation.

\section*{Ethics Statement}
This work focuses on improving the ranking results of passage retrieval systems. Retrieval is the fundamental component for many downstream tasks. However, it poses risks in terms of bias, misuse and misinformation due to the yet inaccurate results. Selection bias resulting from data collection, \textit{e.g.}, lexical bias, may exist in the adopted datasets. Additionally, as the reranking approach in this work is built upon off-the-shelf retrievers, bias may ensue from upstream retrievers. 

\section*{Acknowledgements}
This work was supported by NSFC under Contract U20A20183 and 62021001.

% Entries for the entire Anthology, followed by custom entries
\bibliography{anthology,custom}
\bibliographystyle{acl_natbib}

\clearpage
\appendix

\section{Datasets Details}
\label{appdx:dataset_details}

\begin{table*}[t]
	\centering
	\small
	\begin{tabular}{lcccc}
		\toprule
		& Natural Questions & MS MARCO & TREC-DL 2019 & TREC-DL 2020  \\
		\midrule
		\# Passages in Corpus & 20,015,324 & 8,841,823 & - & - \\
		Avg. Passage Length & 100.0 & 56.58 & - & - \\
		Avg. Query Length & 9.20 & 5.97 & - & - \\
		\# Train Queries & 58,880 & 502,939 & - & - \\
		\# Dev Queries & 6,515 & 6,980 & - & - \\
		\# Test Queries & 3,610 & - & 43 & 54 \\
		\# Train Pairs & 498,816 & 532,761 & - & - \\
		\# Dev Pairs & 55,121 & 7,437 & - & - \\
		\# Test Pairs & - & - & 9,260 & 11,386 \\
		\bottomrule
	\end{tabular}
	\caption{Statistics of Natural Questions, MS MARCO and TREC 2019/2020 datasets.}
	\label{tab:dataset_stat}
\end{table*}

Dataset Natural Questions is under CC BY-SA 3.0 license. MS MARCO and TREC 2019/2020 are under CC BY-SA 4.0 license. The statistics of these datasets are presented in Table~\ref{tab:dataset_stat}.

\section{Full Evaluation Results}
\label{appdx:full_results}
We present the full evaluation results on Natural Questions, MS MARCO and TREC 2019/ 2020 in Table~\ref{tab:full_results_NQ} and \ref{tab:full_results_MSMARCO}.
\begin{table*}[h]
	\centering
	\small
	\begin{tabular}{llllll}
		\toprule
		& \multicolumn{5}{c}{\textbf{Natural Questions Test}} \\
		\cmidrule(lr){2-6}
		& \multicolumn{1}{c}{R@1} & \multicolumn{1}{c}{R@5} & \multicolumn{1}{c}{R@10} & \multicolumn{1}{c}{R@20} & \multicolumn{1}{c}{R@50} \\
		\midrule
		DPR-Multi & 45.82 & 68.12 & 75.23 & 80.31 & 84.57 \\
		DPR-Multi + HybRank & 51.99 (\textbf{+6.17}) & 72.71 (\textbf{+4.59}) & 79.03 (\textbf{+3.80}) & 83.24 (\textbf{+2.93}) & 85.93 (\textbf{+1.36}) \\
		\midrule
		DPR-Single & 47.95 & 69.39 & 75.93 & 80.97 & 84.90 \\
		DPR-Single + HybRank & 53.13 (\textbf{+5.18}) & 73.05 (\textbf{+3.66}) & 78.84 (\textbf{+2.91}) & 82.99 (\textbf{+2.02}) & 85.93 (\textbf{+1.03}) \\
		\midrule
		FiD-KD & 50.36 & 74.10 & 79.78 & 84.27 & 87.90 \\
		FiD-KD + HybRank & 52.85 (\textbf{+2.49}) & 74.46 (\textbf{+0.36}) & 80.50 (\textbf{+0.72}) & 84.49 (\textbf{+0.22}) & 88.06 (\textbf{+0.16}) \\
		\midrule
		ANCE & 52.66 & 72.66 & 78.70 & 83.05 & 86.29 \\
		ANCE + HybRank & 53.63 (\textbf{+0.97}) & 73.57 (\textbf{+0.91}) & 79.28 (\textbf{+0.58}) & 83.88 (\textbf{+0.83}) & 87.12 (\textbf{+0.83}) \\
		\midrule
		RocketQA-retriever & 51.74 & 74.02 & 80.00 & 83.99 & 87.34 \\
		RocketQA-retriever + HybRank & 56.07 (\textbf{+4.33}) & 77.04 (\textbf{+3.02}) & 82.30 (\textbf{+2.30}) & 85.68 (\textbf{+1.69}) & 88.17 (\textbf{+0.83}) \\
		\midrule
		RocketQA-reranker & 54.60 & 76.59 & 81.44 & 85.01 & 88.17 \\
		RocketQA-reranker + HybRank & 59.83 (\textbf{+5.23}) & 78.73 (\textbf{+2.14}) & 82.83 (\textbf{+1.39}) & 86.40 (\textbf{+1.39}) & 88.42 (\textbf{+0.25}) \\
		\midrule
		RocketQAv2-retriever & 55.57 & 75.98 & 81.08 & 84.46 & 87.92 \\
		RocketQAv2-retriever + HybRank & 56.98 (\textbf{+1.41}) & 76.65 (\textbf{+0.67}) & 81.94 (\textbf{+0.86}) & 85.76 (\textbf{+1.30}) & 88.61 (\textbf{+0.69}) \\
		\midrule
		RocketQAv2-reranker & 57.17 & 75.98 & 81.00 & 84.71 & 87.92 \\
		RocketQAv2-reranker + HybRank & 59.50 (\textbf{+2.33}) & 78.34 (\textbf{+2.36}) & 83.24 (\textbf{+2.24}) & 86.26 (\textbf{+1.55}) & 88.75 (\textbf{+0.83}) \\
		\bottomrule
	\end{tabular}
	\caption{The full evaluation of reranking results from HybRank on Natural Questions. We build HybRank upon DPR-Multi~\citep{karpukhin-etal-2020-dense}, DPR-Single~\citep{karpukhin-etal-2020-dense}, FiD-KD~\citep{Izacard2021DISTILLING}, ANCE~\citep{Xiong2021APPROXIMATE}, the retriever and reranker of RocketQA~\citep{qu-etal-2021-rocketqa} and RocketQAv2~\citep{ren-etal-2021-rocketqav2}. Improvements brought by HybRank are highlighted in bold.}
	\label{tab:full_results_NQ}
\end{table*}

\begin{table*}[h]
	\centering
	\small
	\begin{tabular}{llllll}
		\toprule
		& \multicolumn{3}{c}{\textbf{MS MARCO Dev}} & \multicolumn{1}{c}{\textbf{TREC 2019}} & \multicolumn{1}{c}{\textbf{TREC 2020}} \\
		\cmidrule(lr){2-4} \cmidrule(lr){5-5} \cmidrule(lr){6-6}
		\multirow{2}{*}{} & \multicolumn{1}{c}{MRR@10} & \multicolumn{1}{c}{R@10} & \multicolumn{1}{c}{R@50} & \multicolumn{1}{c}{NDCG@10} & \multicolumn{1}{c}{NDCG@10} \\
		\midrule
		DistilBERT-KD & 32.50 & 58.77 & 79.24 & 69.23 & 60.58 \\
		DistilBERT-KD + HybRank & 36.24 (\textbf{+3.74}) & 64.40 (\textbf{+5.63}) & 82.02 (\textbf{+2.78}) & 72.55 (\textbf{+3.32}) & 66.71 (\textbf{+6.13}) \\
		\midrule
		ANCE & 33.01 & 59.44 & 80.10 & 62.37 & 60.00 \\
		ANCE + HybRank & 36.44 (\textbf{+3.43}) & 64.63 (\textbf{+5.19}) & 82.79 (\textbf{+2.69}) & 70.41 (\textbf{+8.04}) & 63.70 (\textbf{+3.70}) \\
		\midrule
		TCT-ColBERT-v1 & 33.49 & 60.46 & 80.67 & 65.42 & 61.03 \\
		TCT-ColBERT-v1 + HybRank & 36.23 (\textbf{+2.74}) & 64.96 (\textbf{+4.50}) & 83.44 (\textbf{+2.77}) & 73.21 (\textbf{+7.79}) & 66.91 (\textbf{+5.88}) \\
		\midrule
		TAS-B & 34.44 & 62.94 & 83.44 & 70.49 & 63.89 \\
		TAS-B + HybRank & 36.38 (\textbf{+1.94}) & 65.77 (\textbf{+2.83}) & 84.71 (\textbf{+1.27}) & 74.82 (\textbf{+4.33}) & 66.53 (\textbf{+2.64}) \\
		\midrule
		TCT-ColBERT-v2 & 35.85 & 63.64 & 83.31 & 71.15 & 64.32 \\
		TCT-ColBERT-v2 + HybRank & 37.55 (\textbf{+1.70}) & 66.39 (\textbf{+2.75}) & 84.97 (\textbf{+1.66}) & 74.06 (\textbf{+2.91}) & 66.35 (\textbf{+2.03}) \\
		\midrule
		RocketQA-retriever & 35.77 & 64.01 & 83.41 & 70.49 & 63.74 \\
		RocketQA-retriever + HybRank & 36.97 (\textbf{+1.20}) & 65.67 (\textbf{+1.66}) & 84.91 (\textbf{+1.50}) & 74.79 (\textbf{+4.30}) & 67.25 (\textbf{+3.51}) \\
		\midrule
		RocketQA-reranker & 40.51 & 69.81 & 86.46 & 75.40 & 67.66 \\
		RocketQA-reranker + HybRank & 40.98 (\textbf{+0.47}) & 70.40 (\textbf{+0.59}) & 86.55 (\textbf{+0.09}) & 77.05 (\textbf{+1.65}) & 69.85 (\textbf{+2.19}) \\
		\midrule
		RocketQAv2-retriever & 37.28 & 65.72 & 84.04 & 70.14 & 63.04 \\
		RocketQAv2-retriever + HybRank & 38.74 (\textbf{+1.46}) & 68.12 (\textbf{+2.40}) & 85.96 (\textbf{+1.92}) & 73.63 (\textbf{+3.49}) & 67.87 (\textbf{+4.83}) \\
		\midrule
		RocketQAv2-reranker & 41.15 & 69.99 & 86.55 & 73.24 & 69.47 \\
		RocketQAv2-reranker + HybRank & 41.40 (\textbf{+0.25}) & 70.37 (\textbf{+0.38}) & 86.68 (\textbf{+0.13}) & 74.92 (\textbf{+1.68}) & 70.71 (\textbf{+1.24}) \\
		\bottomrule
	\end{tabular}
	\caption{The full evaluation of reranking results from HybRank on MS MARCO and TREC 2019/2020. We built HybRank upon DistilBERT-KD~\citep{Hofstatter2021Improving}, ANCE~\citep{Xiong2021APPROXIMATE}, TCT-ColBERT-v1~\citep{Lin2020Distilling}, TAS-B~\citep{Hofstatter2021Efficiently}, TCT-ColBERT-v2~\citep{lin-etal-2021-batch}, the retriever and reranker of RocketQA~\citep{qu-etal-2021-rocketqa} and RocketQAv2~\citep{ren-etal-2021-rocketqav2}. Improvements brought by HybRank are highlighted in bold.}
	\label{tab:full_results_MSMARCO}
\end{table*}

\section{Reranking Cases}
\label{appdx:examples}
\begin{figure*}[h]
	\centering
	\small
	\begin{tabular}{m{0.78\textwidth}<{\raggedright}m{0.16\textwidth}<{\centering}}
		\toprule
		\multicolumn{2}{c}{\textbf{Query: \textit{Who was the king of England in 1756?}}} \\
		
		\toprule
		\multicolumn{2}{c}{\includegraphics[width=1\textwidth]{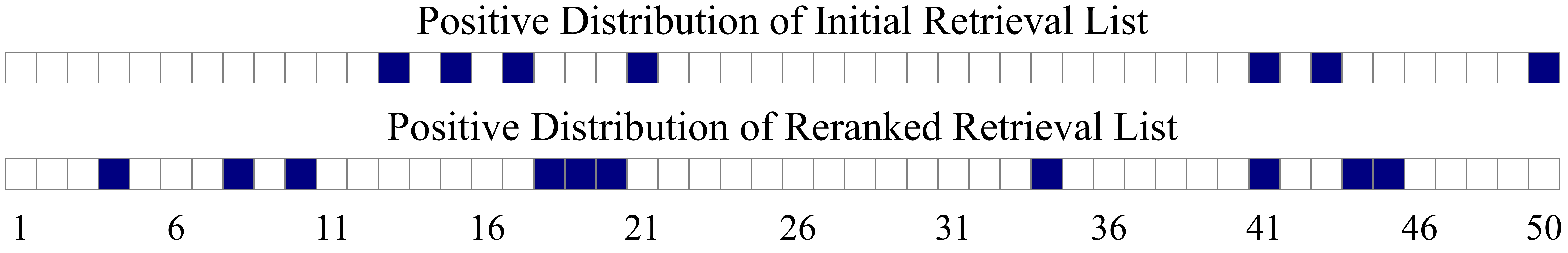}} \\
		\toprule
		
		\multicolumn{1}{c}{Positive Passages} & Rank Changes \\
		\midrule
		\textbf{George II of Great Britain.} \textblue{George II} of Great Britain \textblue{George II} (George Augustus; ; 30 October / 9 November 1683 – 25 October 1760) was King of Great Britain and Ireland, Duke of Brunswick-Lüneburg (Hanover) and a prince-elector of the Holy Roman Empire from 11 June 1727 (O.S.) until his death in 1760. George was the last British monarch born outside Great Britain: he was born and brought up in northern Germany. His grandmother, Sophia of Hanover, became second in line to the British throne after about 50 Catholics higher in line were excluded by the Act of Settlement 1701 and the Acts of & 15 $\rightarrow$ 4 (11 $\uparrow$) \\
		\midrule
		\textbf{George II of Great Britain.} by his grandson, George III. For two centuries after \textblue{George II}'s death, history tended to view him with disdain, concentrating on his mistresses, short temper, and boorishness. Since then, most scholars have reassessed his legacy and conclude that he held and exercised influence in foreign policy and military appointments. George was born in the city of Hanover in Germany, and was the son of George Louis, Hereditary Prince of Brunswick-Lüneburg (later King George I of Great Britain), and his wife, Sophia Dorothea of Celle. His sister, Sophia Dorothea, was born when he was three years old. Both of George's parents & 74 $\rightarrow$ 8 (66 $\uparrow$)  \\
		\midrule
		\textbf{Monarchy of the United Kingdom.} Britain was now in personal union. Power shifted towards George's ministers, especially to Sir Robert Walpole, who is often considered the first British prime minister, although the title was not then in use. The next monarch, \textblue{George II}, witnessed the final end of the Jacobite threat in 1746, when the Catholic Stuarts were completely defeated. During the long reign of his grandson, George III, Britain's American colonies were lost, the former colonies having formed the United States of America, but British influence elsewhere in the world continued to grow, and the United Kingdom of Great Britain and Ireland was created & 17 $\rightarrow$ 10 (7 $\uparrow$) \\
		\midrule
		\textbf{Duke of Cumberland.} of Wales, the eldest son and heir apparent of King \textblue{George II} and the father of King George III. He died without legitimate issue, when the dukedom again became extinct. This double dukedom, in the Peerage of Great Britain, was bestowed on Prince Ernest Augustus (1771–1851) (later King of Hanover), the fifth son and eighth child of King George III of the United Kingdom and King of Hanover. In 1919 it was suspended under the Titles Deprivation Act 1917 and has not been restored to its titular heir. A historic fixed bridge hand is known as the Duke of Cumberland & 67 $\rightarrow$ 18 (49 $\uparrow$) \\
		\midrule
		\textbf{George II of Great Britain.} the Hanoverian quarter differenced overall by a label of three points argent. The crest included the single arched coronet of his rank. As king, he used the royal arms as used by his father undifferenced. Caroline's ten pregnancies resulted in eight live births. One of their children died in infancy, and seven lived to adulthood. \textblue{George II} of Great Britain \textblue{George II} (George Augustus; ; 30 October / 9 November 1683 – 25 October 1760) was King of Great Britain and Ireland, Duke of Brunswick-Lüneburg (Hanover) and a prince-elector of the Holy Roman Empire from 11 June 1727 (O.S.) until & 13 $\rightarrow$ 19 (6 $\downarrow$)  \\
		\bottomrule
	\end{tabular}
	\caption{Reranking case 1. Blue squares indicate positive passages and white squares stand for negative passages. The titles of passages are bold and put in front of passages. These blue texts are the answers for the question.}
	\label{fig:rerank_case_1}
\end{figure*}

\begin{figure*}[h]
	\centering
	\small
	\begin{tabular}{m{0.78\textwidth}<{\raggedright}m{0.16\textwidth}<{\centering}}
		\toprule
		\multicolumn{2}{c}{\textbf{Query: \textit{What kind of book is Where's Waldo?}}} \\
		
		\toprule
		\multicolumn{2}{c}{\includegraphics[width=1\textwidth]{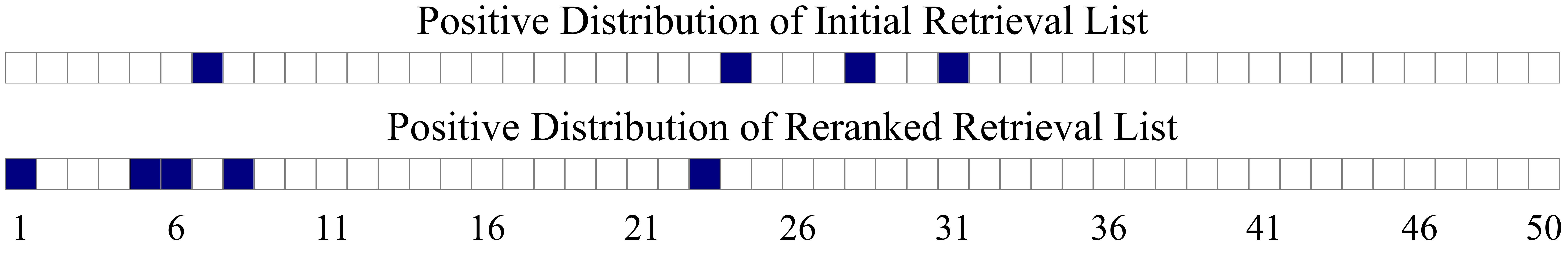}} \\
		\toprule
		
		\multicolumn{1}{c}{Positive Passages} & Rank Changes \\
		\midrule
		\textbf{Where's Waldo? (video game).} Where's Waldo? (video game) Where's Waldo? (Where's Wally? in the UK, Australia and South America) is a \textblue{puzzle} game developed by Bethesda Softworks and published in 1991 by THQ for the Nintendo Entertainment System. It was the first video game loosely based on Martin Handford's book of the ``same name''. Mostly similar to the books, players must help Waldo get to the moon by finding him in each of the eight levels in the game. The game was panned by critics, who criticized the game for its graphics, which made it more difficult to find Waldo in each of the & 24 $\rightarrow$ 1 (23 $\uparrow$) \\
		\midrule
		\textbf{Where's Waldo? (video game).} takes advantage of superior pointer-based motion controls to easily locate search targets and supports versus multiplayer. Where's Waldo? (video game) Where's Waldo? (Where's Wally? in the UK, Australia and South America) is a \textblue{puzzle} game developed by Bethesda Softworks and published in 1991 by THQ for the Nintendo Entertainment System. It was the first video game loosely based on Martin Handford's book of the ``same name''. Mostly similar to the books, players must help Waldo get to the moon by finding him in each of the eight levels in the game. The game was panned by critics, who criticized the & 31 $\rightarrow$ 5 (26 $\uparrow$)  \\
		\midrule
		\textbf{Activity book.} and does not fall neatly into one of these more specific categories. Activity books are typically centred around a particular theme. This may be a generic theme, e.g. dinosaurs, or based on a toy, television show, book, or game. For example, the Where's Wally? series of books (known as Where's Waldo? in the USA) by Martin Handford consists of both \textblue{puzzle} books, wherein the reader must search for characters hidden in pictures, and activity books such as ``'', which include a wider range of games and activities as well as \textblue{puzzles}. In 2018, Nintendo announced its intention to publish activity & 28 $\rightarrow$ 6 (22 $\uparrow$) \\
		\midrule
		\textbf{Where's Waldo? The Fantastic Journey (video game).} Where's Waldo? The Fantastic Journey (video game) Where's Waldo? The Fantastic Journey is a video game published by Ubisoft and developed by Ludia based on the book of the same name. It is a \textblue{puzzle} adventure game released for the Nintendo DS, Wii, Microsoft Windows, and the iPhone, and is also a remake of ``The Great Waldo Search'', released in 1992. Like the other games in the series, the object of the game is to search for hidden characters and items within a time limit. Hints are awarded to the player through Woof, Waldo's pet dog. Woof alerts the players & 7 $\rightarrow$ 8 (1 $\downarrow$) \\
		\midrule
		\textbf{Where's Wally?} was turned into a Sunday newspaper comic/\textblue{puzzle}, distributed by King Features Syndicate. The comics were also released in book form in the US, using the regional name `Waldo'. In the early 1990s Quaker Life Cereal in the US carried various ``Where's Wally?'' scenes on the back of the boxes along with collector's cards, toys and send-away prizes. This was shown in ``The Simpsons'' episode ``Hello Gutter, Hello Fadder'' where Homer shouts ``WALDO, WHERE ARE YOU?!'' after looking at the scene on the cereal box as Waldo walks by the kitchen window. On 1 April 2018 Google Maps added a minigame & 61 $\rightarrow$ 23 (38 $\uparrow$)  \\
		\bottomrule
	\end{tabular}
	\caption{Reranking case 2. Blue squares indicate positive passages and white squares stand for negative passages. The titles of passages are bold and put in front of passages. These blue texts are the answers for the question.}
	\label{fig:rerank_case_2}
\end{figure*}

We present reranking cases in Figure~\ref{fig:rerank_case_1} and Figure~\ref{fig:rerank_case_2}. The first lines in these figures are the query sentence. We illustrate the distribution of positives in the passage list before and after reranking. Blue squares indicate positive passages while white squares stand for negative passages in the retrieval list. We only show top-$50$ out of $100$ passages in these lists due to the space limitation. Following the positive distribution, we list several raw texts of reranked passages for the question.

Observed from the distribution visualization and rank changes of passages, the positive distributions shift toward the front of the lists as the quantitative analysis in Section~\ref{sec:results}. Ranks of many positive passages are raised by a large margin. Besides, it is apparent that positive passages tend to describe the same entities, events and relations as discussed in Section~\ref{sec:Introduction}. Case 1 in Figure~\ref{fig:rerank_case_1} involves ``the king of England'' while case 2 in Figure~\ref{fig:rerank_case_2} is about ``Where's Waldo''.

\end{document}